# Biases from Missing a Small Planet in High Multiplicity Systems

C. Alexander Thomas[1], Lauren M. Weiss[1], and Matthias Y. He[2,3]
[1] Department of Physics and Astronomy, University of Notre Dame, Notre Dame, IN 46556, USA
[2] NASA Ames Research Center, Moffett Field, CA 94035, USA

## Abstract

In an era when we are charting multiple planets per system, one might wonder the extent to which "missing" (or failing to detect) a planet can skew our interpretation of the system architecture. We address this question with a simple experiment: starting from a large, homogeneous catalog, we remove planets and monitor how several well-defined metrics of the system architecture change. We first perform this test on a catalog of observed exoplanets. We then repeat our test on a catalog of synthetic planetary systems with underlying hyperparameters that have been fit to reproduce the observed systems as faithfully as possible (though imperfectly). For both samples, we find that the failure to detect one or more planets tends to create more irregularly spaced planets, whereas the planet mass similarity and coplanarity are essentially unaffected. One key difference between the synthetic and observed data sets is that the observed systems have more evenly spaced planets than the observation-bias-applied synthetic systems. Since our tests show that detection bias tends to increase irregularity in spacing, the even spacing in the observed planetary systems is likely astrophysical rather than the result of the Kepler missions' inherent detection biases. Our findings support the interpretation that planets in the same system have similar sizes and regular spacing and reinforce the need to develop an underlying model of planetary architectures that reproduces these observed patterns.

*Unified Astronomy Thesaurus concepts:* Exoplanet astronomy (486); Exoplanets (498)

## 1. Introduction

The NASA Kepler Mission is responsible for the detection of over 4000 planet candidates, of which many are arranged in systems with multiple planets. Based on these detections, it is likely that 30%–80% of the ∼190,000 stars in the Kepler Mission host multiple planets (G. D. Mulders et al. 2018; W. Zhu et al. 2018). One of the key planetary structures is the "peas-in-a-pod" pattern, which describes planetary systems with uniformly sized and spaced planets (L. M. Weiss et al. 2018). Although the nature of this pattern has been questioned (L. Murchikova & S. Tremaine 2020; W. Zhu 2020), numerous studies have found evidence supporting the astrophysical origin of peas-in-a-pod (S. Millholland et al. 2017; G. D. Mulders et al. 2018; L. M. Weiss et al. 2018; M. Y. He et al. 2019; E. Sandford et al. 2019; G. J. Gilbert & D. C. Fabrycky 2020; L. M. Weiss & E. A. Petigura 2020). The uniformity also extends to planet masses (S. Millholland et al. 2017; S. Wang 2017) and is present for systems of planets that are both rocky and gas-enveloped (S. C. Millholland & J. N. Winn 2021). The pattern appears to have an edge at about 100–300 days (S. C. Millholland et al. 2022), beyond which additional transiting planets fitting the pattern were detectable but not found. In addition, rocky systems are less uniform in mass yet more uniform in size and spacing than systems containing volatile-rich planets (A. V. Goyal & S. Wang 2024). These mounting lines of evidence suggest that the peas-in-a-pod pattern is a fossil that, if successfully decoded, could reveal one of the most common modes of planet formation in the galaxy.

That mode of planetary formation is the subject of ongoing theoretical investigations (F. C. Adams 2019; F. C. Adams et al. 2020; K. Batygin & A. Morbidelli 2023).

However, one major setback in investigating the peas-in-a-pod pattern is that, for most systems, there is no guarantee that we have detected all of the planets. In particular, nondetections could result in underlying architectures that differ, perhaps substantially, from the architectures we infer from the observed planets. For example, the presence of an outer giant may affect the observed orbital spacing of a planetary system (M. Y. He & L. M. Weiss 2023; L. M. Weiss et al. 2024; J. R. Livesey & J. Becker 2025). The central question of this Letter is: how does the failure to detect a planet in compact multiplanet systems affect our interpretation of planetary system architecture?

To address the above question, we perform a simple experiment that involves removing one planet at a time from compact multiplanet systems (i.e., "jackknife" testing; W. H. Press et al. 2007) to measure the extent to which the removal of a planet affects our interpretation of the system. We also performed a similar experiment using synthetic underlying catalogs of planetary systems generated by SysSim (M. Y. He et al. 2019, 2020), comparing their underlying architectures to the subset of planets that would be detected after accounting for detection biases. In particular, we address: how does removing (i.e., failing to detect) one or more planets affect our interpretation(s) of (1) the uniform sizes (or masses) of the planets, (2) the uniform spacing of the planets, and (3) the flatness, or coplanarity, of the planetary system. Based on the observation that many compact multiplanet systems appear to have the peas-in-a-pod pattern, we expect: (1) the size and/or mass uniformity should be unbiased by a missed planet, (2) the uniform spacing should be disrupted if a "middle" planet (neither the innermost nor outermost) is missed, whereas the uniform spacing should not be disrupted if an edge planet is missed, and (3) the measure of flatness should be unaffected by missing a planet (assuming the missed planet is itself transiting).

---

[3] NASA Postdoctoral Program Fellow.

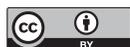

To answer these questions, we quantify the mass, coplanarity, and spacing distributions of planets within a system. Two of the metrics that we use are gap complexity and mass partitioning. Gap complexity is defined as

$$\mathcal{C} = -K \left( \sum_{k=1}^{n} p_k^* \log p_k^* \right) \cdot \left( \sum_{k=1}^{n} \left( p_k^* - \frac{1}{n} \right)^2 \right), \quad (1)$$

where $K$ is a normalization constant, $n = N - 1$ is the number of gaps between $N$ planets, and $p_k^*$ are defined as the "pseudo-probabilities" and are normalized to one:

$$p_k^* \equiv \frac{\log(P_{k+1}/P_k)}{\log(P_{\max}/P_{\min})}, \quad (2)$$

where $P_{\max}$ and $P_{\min}$ are, respectively, defined as the maximum period and the minimum period of the known planets in the system while $P_k$ and $P_{k+1}$ are the inner and outer periods for the $k$th pair of adjacent planets. Planets are indexed from 0 to N, from the most interior planet to the most exterior planet. The range of values for gap complexity is zero (evenly spaced planets) to one (maximum $p_k^* \approx 2/3$ G. J. Gilbert & D. C. Fabrycky 2020).

Similarly, mass partitioning, which describes the dissimilarity of planet masses in a single system, is given by

$$\mathcal{Q} \equiv \left( \frac{N}{N-1} \right) \cdot \left( \sum_{k=1}^{N} \left( m_k^* - \frac{1}{N} \right)^2 \right), \quad (3)$$

where

$$m_k^* = m_k / \sum_{k}^{N} m_k, \quad (4)$$

where $m_k$ is the mass of the $k$th planet (G. J. Gilbert & D. C. Fabrycky 2020). Unlike gap complexity, the order of the planets does not matter for mass partitioning. The normalizing coefficient, $N/(N-1)$, restricts $\mathcal{Q}$ to be from 0 to 1, where a value of 0 indicates equal masses throughout the system, whereas $\mathcal{Q} \rightarrow 1$ represents a system in which one planet has almost all of the mass and the remaining $N - 1$ planets have vanishingly small masses.

In order to characterize the coplanarity of the systems in our samples, we calculated the dispersion of the impact parameters. The impact parameter is a degenerate parameter, which may lead to different configurations with the exact same dispersion of impact parameters. We use the impact parameter dispersion as a proxy for the coplanarity of the system. As one mechanism of missing planets is due to an inclined orbit, understanding how the inclination of the planets within a system are distributed is important. The impact parameter is

$$b_k = (a_k \cos(i_k))/R_\star,$$

where $a_k$ is the planet orbital distance of the $k$th planet, $R_\star$ is the stellar radius, and $i_k$ is the inclination between the planet orbital plane and the sky plane (transiting configurations have $i \approx 90°$). We calculate the dispersion of impact parameters by taking the standard deviation within a system:

$$\sigma_b = \sqrt{\frac{\Sigma_k (b_k - \bar{b})^2}{N}}, \quad (5)$$

where $N$ is the number of planets in the planetary system. We adopt this metric because the impact parameter can be computed for nontransiting planets ($b > 1$), allowing us to compare this estimation of the coplanarity of systems before and after planets are removed.

In addition to being able to quantify the characteristics of individual planetary systems, these metrics can be used to perform analysis across many different systems. For example, M. Y. He & L. M. Weiss (2023) showed that of the systems with three transiting planets, those that had high gap complexity ($\mathcal{C} > 0.3$) were more likely to have a Jovian planet orbiting within 5 AU than those with more regular spacing. This provides evidence that there may be a link between the gap complexity of the known planets in a multiplanet system and the potential presence of an additional, undetected planet.

Additionally, super-Earths and sub-Neptunes have been discovered in apparent gaps in the Kepler multiplanet systems. Radial velocities led to the discovery of Kepler-20 g, a nontransiting planet in a gap in a system with five transiting planets (L. A. Buchhave et al. 2016). Reanalysis of Kepler photometry has also yielded the detection of transiting planets and/or candidates in what first appeared to be gaps, including the eighth planet in Kepler-90 (C. J. Shallue & A. Vanderburg 2018) and the fifth, sixth, and seventh planet candidates in Kepler-385 (J. J. Lissauer et al. 2024). There are systems, such as Kepler-219 and Kepler-286, that contain 3+ transiting planets of similar size yet have high gap complexity ($\mathcal{C} = 0.72$ and $\mathcal{C} = 0.40$, respectively). These systems are above the gap complexity threshold described in M. Y. He & L. M. Weiss (2023), yet do not have a detected outer giant in the system. One explanation for the irregularity in spacing compared to the regularity in sizes is that these systems host additional small planets. With this possibility in mind, we investigate how metrics change for systems missing planets.

## 2. Methods and Results

### 2.1. Astrophysical Sample

The first sample we analyze is J. J. Lissauer et al. (2024, L24), which contains over 4300 transiting planet candidates identified in the NASA Kepler photometry with homogeneously determined orbital periods and transit depths. We removed any planets with a single transit, denoted by a negative period in L24. In addition, we only used the planets that are identified as a planet by the disposition scheme that L24 defines, which counts both confirmed planets and planet candidates as bona fide. Finally, as gap complexity can only be calculated for systems with at least three planets, we removed any planets that are in a system with two or fewer planets. These data cuts result in a data set of 873 planets within 252 systems. We will refer to this sample as the *Astrophysical—Observed* sample. Note that, for systems that contain additional, nontransiting planets, such as Kepler-20, we do not manually add those planets to the L24 catalog. Many of the planets in L24 do not have measured masses. To explore the possible diversity of planet masses at a given radius, both within a given system and throughout our sample, we used the Forecaster probabilistic mass–radius relationship program from

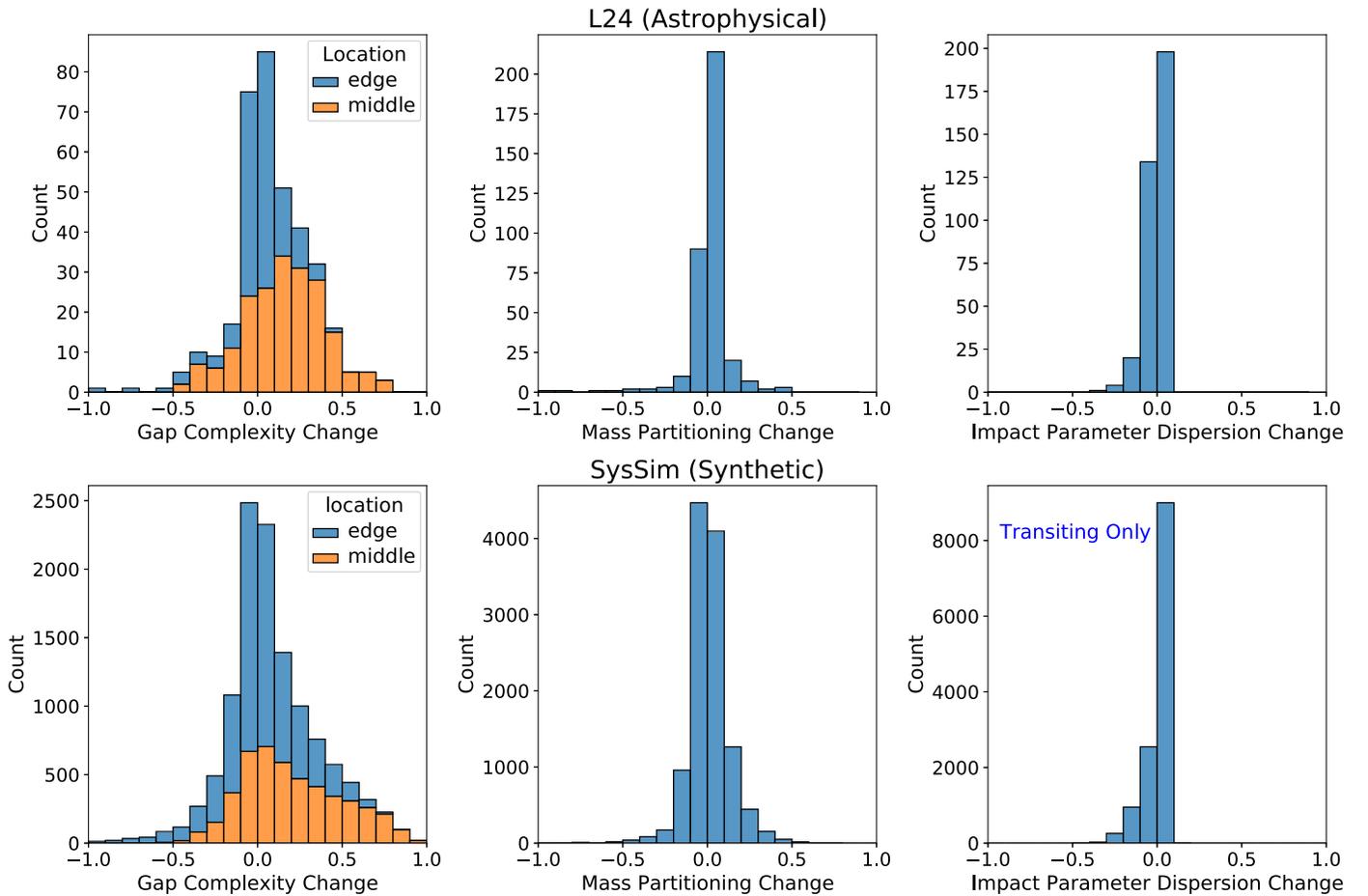

**Figure 1.** Top: histograms of the difference in gap complexity (left), mass partitioning (center), and impact parameter dispersion (right) of individual systems in L24 once a planet has been removed. The gap complexity sample is a stacked histogram of cases when an edge planet is removed (blue) and when a middle planet is removed (orange). Bottom: the same as the top for the change in system metrics between SysSim's synthetic underlying and detected samples. The gap complexity sample is now split between cases when all planets removed are edge planets (blue) vs. cases when at least one middle planet is removed (orange). The systems in the bottom right panel have been downselected to include only the transiting planets in the SysSim catalogs.

J. Chen & D. Kipping (2017) to estimate the individual planet masses, sampling each planetary system 100 times. We first calculated the gap complexity, mass partitioning, and impact parameter dispersion for each system in the Astrophysical—Observed data set. To create a comparison sample, we generated systems by removing a single planet from a system in the Astrophysical—Observed data set; this process is done for each planet in a system, in a manner akin to jackknife resampling. This creates the sample that we call the *Astrophysical—One Planet Removed* sample. We must use systems with at least four transiting planets in the Astrophysical—Observed data set, such that we will be able to calculate a gap complexity of 3+ planets after removing one. The Astrophysical Observed sample has 357 planets that are in systems of 4+ transiting planets, in 80 systems. As there are 357 planets in that sample, there are 357 systems in the One Planet Removed sample (one jackknife system for each planet removed).

We computed how each metric changed in each system before and after removing a planet. These distributions are shown in Figure 1 (see also Table 1, fourth column). For gap complexity (upper left panel), we distinguish whether the planet was removed from the edge of the detected system ("edge planet removed," blue) or from the middle ("middle planet removed," orange) in a stacked histogram. For gap complexity, there is a significant change when a planet is removed ($\Delta_{50\%}$ system $= +0.05$). Note that $\Delta_{50\%}$ system is the median difference between individual L24 and L24-1TP systems. The bulk of this change arises when a "middle" planet is removed ($\Delta_{50\%}$ system $= +0.15$) rather than an edge planet ($\Delta_{50\%}$ system $< 0.01$). The mass partitioning (upper middle panel) and the impact parameter dispersion (upper right panel) of each system do not change significantly when a planet is removed ($\Delta_{50\%}$ system $< 0.01$).

We also created cumulative distribution functions (CDFs) for both the Astrophysical–Observed and Astrophysical–One Planet Removed samples, which allowed us to investigate how these metrics change when applied to the overall populations, rather than individual systems. The median values of these distributions, as well as the differences in the medians of these distributions, are also shown in the first three columns of Table 1. Importantly, we were able to perform Kolmogorov–Smirnov (KS) and Anderson–Darling (AD) tests to investigate whether the differences between the distributions were significant. Figure 2 shows the CDFs of gap complexity, mass partitioning, and impact parameter dispersion for the Astrophysical—Observed and the Astrophysical—One Planet Removed samples. For all tests, we adopt $p < 0.02$ as significant, as a balance between our sample sizes and the number of KS and AD tests we performed.

For the gap complexity distributions (Figure 2, top left panel), there is a significant statistical difference between the Astrophysical–Observed sample and the Astrophysical–One

**Table 1**
Changes in Metrics of the Astrophysical–Observed and Astrophysical–One Planet Removed

| Metric | L24 | L24-1TP | $\Delta_{50\%}$ Dist | $\Delta_{50\%}$ System | KS $P$-value | AD $P$-value |
|---|---|---|---|---|---|---|
| Mass Partitioning | $0.10^{+0.18}_{-0.06}$ | $0.11^{+0.42}_{-0.08}$ | +0.01 | +0.00 | 0.65 | 0.47 |
| Impact Parameter Dispersion ($b<1$) | $0.17^{+0.12}_{-0.08}$ | $0.15^{+0.13}_{-0.09}$ | −0.02 | +0.00 | 0.18 | 0.19 |
| Gap Complexity | $0.08^{+0.29}_{-0.06}$ | $0.19^{+0.34}_{-0.16}$ | +0.11 | +0.05 | 0.0004 | 0.0008 |
| Gap Complexity (edge) | ... | $0.07^{+0.38}_{-0.07}$ | −0.00 | +0.00 | 0.31 | 0.26 |
| Gap Complexity (middle) | ... | $0.28^{+0.26}_{-0.17}$ | +0.20 | +0.15 | $<10^{-5}$ | $<10^{-5}$ |

**Note.** Columns are: L24—each metric is applied to the full sample of J. J. Lissauer et al. (2024); L24-1TP—one transiting planet is removed from each system in the L24 sample, iterating through all possible permutations in each system; $\Delta_{50\%}$ dist—the difference between the L24 and L24-1TP cumulative distributions at the 50th percentile; $\Delta_{50\%}$ system—the median difference between individual L24 and L24-1TP systems; KS $P$-value—the resulting Kolmogorov–Smirnov $p$-value statistic for the L24 and L24-1TP distributions; AD $P$-value—the resulting Anderson–Darling $p$-value statistic for the same distributions. The reported uncertainties contain 68% of the distributions.

Planet Removed sample (KS $p = 0.0004$; AD $p = 0.0008$). As expected, the Astrophysical–One Planet Removed sample has a higher median value (0.19) than the Astrophysical–Observed sample (0.08). Also as expected, the subsample of systems where a "middle" planet is removed drives this result $p < 10^{-5}$ (KS and AD), whereas the subsample where an "edge" planet is removed is effectively indistinguishable from the parent population ($p = 0.31$ (KS) and $p = 0.26$ (AD)). The $p$-value results and the median values of these gap complexity distributions are included as columns (5) and (6) in Table 1.

In contrast to gap complexity, mass partitioning (Figure 2, bottom left panel) does not show a statistically significant difference between the two astrophysical samples. The median of the Observed sample is $\mathcal{Q} = 0.10$ and slightly increases to $\mathcal{Q} = 0.11$ for the One Planet Removed sample. Furthermore, the KS-test $p$-value is 0.65 and the AD-test $p$-value is 0.47, so we cannot rule out a common underlying distribution. One key point is that we are using a nondeterministic mass–radius relation to estimate each planet's mass; the mass partitioning distributions may slightly vary each time we perform this experiment. To address this concern, we performed this experiment 100 times. Across 100 trials, the median mass partitioning for the Observed sample is $0.10^{+0.18}_{-0.06}$ while the One Planet Removed sample has a mass partitioning of $0.11^{+0.42}_{-0.08}$. Removing planets from high multiplicity Kepler systems does not result in a large change in the mass partitioning distribution of a system, as we predicted.

Furthermore, the impact parameter dispersion of L24 is also unaffected by removing a transiting planet from a system (Figure 2, bottom right panel). Removing a planet does not significantly flatten or puff up the coplanarity of the system (with a median change of $\Delta_{50\%}$ dist $= -0.02$ and $p = 0.18$ (KS) and $p = 0.19$ (AD)).

One might wonder if there is something special or anomalous about the L24 observed data set. To explore this possibility, we repeated the above analysis on three additional data sets: the KGPS (L. M. Weiss et al. 2024) data set, the California Kepler Survey data set (B. J. Fulton & E. A. Petigura 2018), and a selection of planets from the NASA Exoplanet Archive[4] (NASA Exoplanet Archive 2019). All of these data sets yielded similar qualitative results to what we obtained with L24 (gap complexity increases when a planet is removed; mass partitioning is unaffected; impact parameter dispersion was not computable for some of these data sets). Given the general consistency of these patterns, we adopt the L24 catalog of observed planets in this Letter as it is the largest and most homogeneous.

### 2.2. Synthetic Sample

While removing one planet from observed astrophysical systems provides a useful benchmark for understanding the effect of missing planets, an important caveat of this analysis is that the observed astrophysical systems are not guaranteed to be excellent representations of the underlying astrophysical systems. For instance, the detection biases of the Kepler Mission favor the discovery of planets that are large, close to their star, and transiting ($b < 1$). The underlying systems might contain planets that are smaller, farther out, and at higher impact parameters (especially nontransiting) than the examples in our Astrophysical—Observed catalog, and we would like to understand how failing to detect such planets biases architectural metrics. To this end, we assess: for a reasonable proposed underlying distribution of planets, which planets would be missed in a Kepler-like mission, and how does missing those planets affect the architecture metrics?

A recently proposed underlying distribution for the Kepler planetary architectures was developed using SysSim (M. Y. He et al. 2019, 2020). To generate this model, several parameters (governing the number of planets drawn in each system, their orbital period and radii distributions, and their dynamical separations and properties) were tuned to optimally (using a weighted distance function via approximate Bayesian computation) reproduce many of the observed distributions that capture the orbital and size properties of the Kepler planets, for both observed single- and multitransiting systems (after incorporating a simple model of detection bias). SysSim uses a cluster point process to generate synthetic planets within synthetic planetary systems, which serves as the Synthetic— Underlying catalog. A procedure similar to the Kepler detection pipeline is applied to downselect to a catalog of detected planets. The goal of this process is to find a model for the underlying distribution of planetary systems that, when observed by a Kepler-like mission, can mimic the known Kepler catalog. Note that this may result in more than one planet being removed from a system, in contrast to our earlier experiment. We will refer to this as the *Synthetic—Detected* catalog. To downselect from the Synthetic—Underlying catalog to the Detected catalog, the simulated planetary systems were initialized by drawing their invariable planes isotropically, and then assigning planetary orbital planes with their mutual inclinations relative to that invariable plane.

---

[4] This sample was selected from the NASA Exoplanet Archive Confirmed Planet table on 24/6/2. We removed any planets observed by direct imaging or pulsar timing. We further removed any planet without a mass measurement. Then we selected only the systems with a multiplicity of 4+ with no planets removed in prior steps.

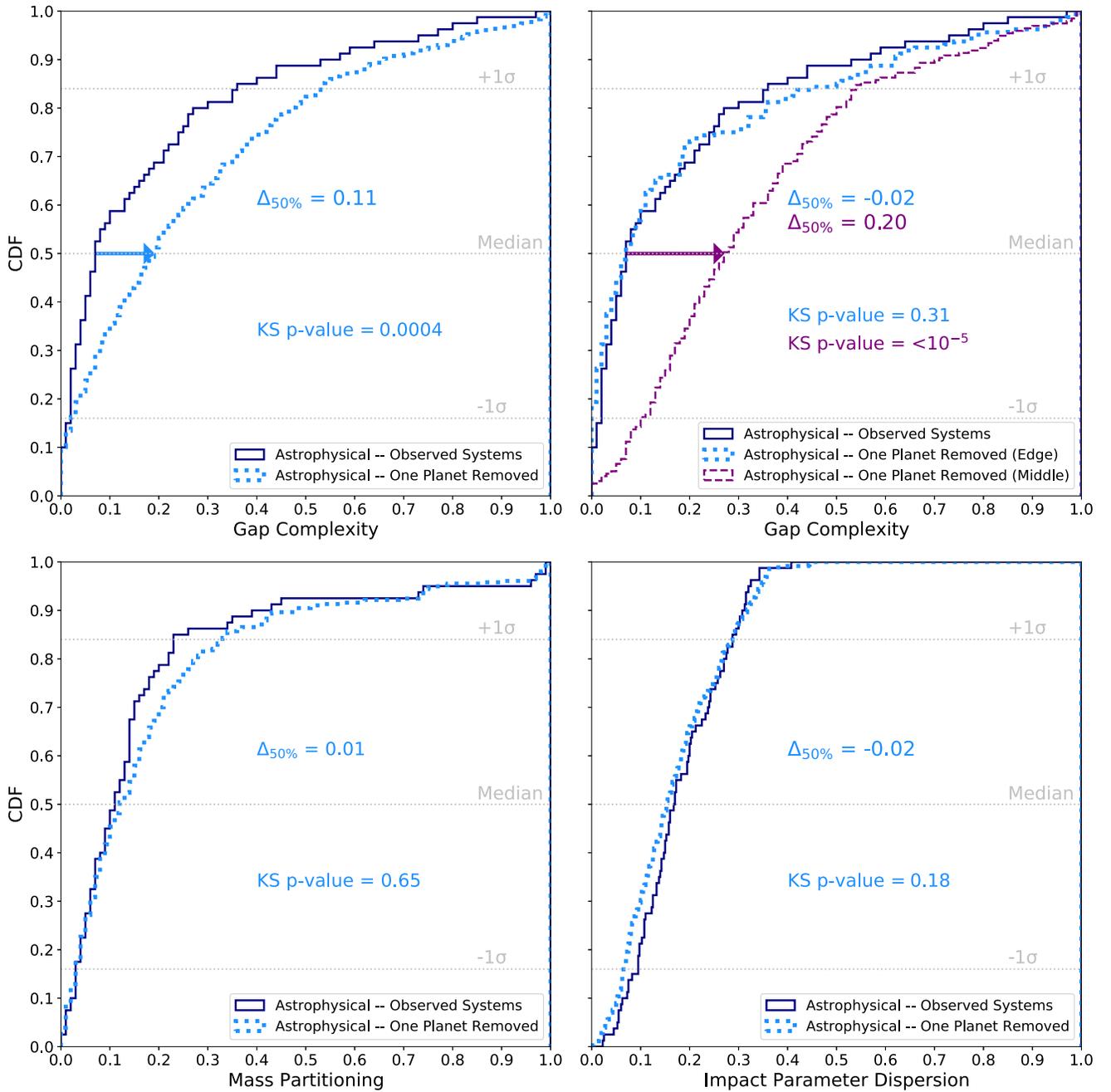

**Figure 2.** Cumulative distribution functions of gap complexity (top), mass partitioning (bottom left), and impact parameter dispersion (bottom right) for a sample of 80 4+ planet L24 systems. The solid lines (dark blue) represent the observed systems, while the dotted lines (light blue) represent the sample of systems with a planet removed. In the top right panel, the One Planet Removed sample is split between systems with an edge planet removed (light blue) and systems with a middle planet removed (purple).

For this Letter, we used the 100 pregenerated catalogs from the "maximum angular momentum deficit (AMD) model" in M. Y. He et al. (2020). We removed systems with fewer than three detected planets (for which gap complexity is not calculable). To reduce the effects of stochastic noise in each pregenerated catalog and to marginalize over the uncertainties in the model parameters, we combined all 100 of the pregenerated `SysSim` synthetic catalogs. This results in 12,786 Synthetic—Detected systems, for which the original corresponding Synthetic—Underlying systems are known.

We calculated the system-wide metrics for each system in the Synthetic—Underlying and Synthetic—Detected samples, and also the change in each metric caused by missing the undetected planets. The bottom row of Figure 1 shows how the architectural metrics in the synthetic systems change due to missing planets (also see Table 2). Overall, there is a positive change in gap complexity ($\Delta_{50\%}$ system $= +0.05$, lower left

**Table 2**
Changes in Metrics of the Synthetic SysSim Maximum-AMD Sample

| Metric | Underlying | Detected | $\Delta_{50\%}$ Dist | $\Delta_{50\%}$ System | KS $p$-value | AD $p$-value |
| --- | --- | --- | --- | --- | --- | --- |
| Mass Partitioning | $0.13^{+0.22}_{-0.07}$ | $0.13^{+0.19}_{-0.09}$ | +0.00 | +0.00 | 0.34 | 0.16 |
| Impact Parameter Dispersion | $0.69^{+0.92}_{-0.44}$ | $0.19^{+0.10}_{-0.09}$ | −0.51 | −0.57 | $<10^{-5}$ | $<10^{-5}$ |
| Impact Parameter Dispersion ($b<1$) | $0.21^{+0.09}_{-0.09}$ | N/A | −0.02 | +0.00 | 0.25 | 0.074 |
| Gap Complexity | $0.13^{+0.18}_{-0.08}$ | $0.18^{+0.39}_{-0.16}$ | +0.05 | +0.05 | 0.0052 | 0.0005 |
| Gap Complexity (edge) | ... | $0.13^{+0.18}_{-0.12}$ | +0.00 | +0.01 | 0.033 | 0.005 |
| Gap Complexity (middle) | ... | $0.29^{+0.43}_{-0.23}$ | +0.16 | +0.16 | 0.00015 | $<10^{-5}$ |

**Note.** Columns are: Underlying—the median values for each metric for 100 pregenerated SysSim underlying catalogs; Detected—the median values for each metric for 100 pregenerated SysSim detected catalogs, which are based on the Underlying catalog; $\Delta_{50\%}$ Dist—the difference between the Underlying and Detected at the 50th percentile; $\Delta_{50\%}$ System—the median difference between individual Underlying and Detected systems; KS $p$-value—the median of the $p$-values between the individual underlying and individual detected catalogs for each metric for 100 pregenerated SysSim physical catalogs; AD $p$-value—same as median KS $p$-value but for a two sample Anderson–Darling test. The reported uncertainties contain 68% of the distributions.

panel) when planets are missed. As before, we split the gap complexity calculations into subsamples; an "edge" sample where all of the missed planets are at the edge(s) of the detected planets, and a "middle" sample where at least one missed planet is in between detected planets. Much like in the L24 data, missing at least one middle planet drives a significant increase in the gap complexity ($\Delta_{50\%}$ system = +0.16), whereas missing planets at the edge(s) of systems have a negligible effect ($\Delta_{50\%}$ system = +0.01). There is no apparent change in the mass partitioning ($\Delta_{50\%}$ system < 0.01, middle panel). An interesting feature of the synthetic sample is that it contains planets that are not transiting ($b>1$), a population that is absent from the L24 sample. If we compare the impact parameter dispersion of *all* the planets (including nontransiting) between the underlying and detected samples, there is a major change in the impact parameter dispersion, due to missing so many nontransiting planets, which by definition have values of $b>1$ ($\Delta_{50\%}$ system = −0.57). However, when we select the subset of underlying planets that are transiting ($b<1$) and compare their systems' impact parameter distributions before and after detection biases are applied, the change in impact parameter dispersion disappears ($\Delta_{50\%}$ system = 0.00). In other words, the transiting planets that are missed (due to their small radii and long orbital periods) are not at preferentially high or low inclinations compared to the detected transiting planets.

The CDFs of the population distributions for the SysSim systems are shown in Figure 3, and their medians and 68% intervals are presented in Table 2. As this combined SysSim catalog contains over 12,000 systems, any slight deviation in the distributions will appear significant in a KS or an AD test. To address this, we split the combined catalog back into the original 100 individual SysSim catalogs. For each catalog, we calculate the metrics and their corresponding KS-test and AD-test $p$-values. We then determined the median $p$-values, which can be seen in Table 2. The differences in the mass partitioning distributions are, in general, not statistically significant ($p = 0.34$ (KS) and $p = 0.16$ (AD)). The impact parameter dispersion shows, strongly, that the synthetic underlying and detected catalogs are statistically different with KS and AD $p$-values $< 10^{-5}$ for every catalog. However, when downselecting for just the transiting planets, the KS and AD $p$-values (0.25 and 0.074, respectively) show that these distributions may be drawn from the same population. Finally, there is a statistically significant difference between the gap complexity distributions of these two samples with $p = 0.0052$ (KS) and $p = 0.0005$ (AD). The difference between the two samples is driven by systems where at least one planet is removed from the middle ($p = 0.00015$ (KS)), as removing only edge planets does not have as significant an impact ($p = 0.033$ (KS)). These results mirror the results for the L24 data set.

### 2.3. Comparison of Astrophysical and Synthetic Catalogs

When comparing the L24 results to the SysSim results, it is important to note that while SysSim is designed to reproduce the observed Kepler catalog, it is not a one-to-one comparison. L24 is based on Kepler Data Release 25 supplemental (DR25supp) catalog and includes Kepler Objects of Interest (KOI) from various other sources (such as R. Sanchis-Ojeda et al. 2014; C. J. Shallue & A. Vanderburg 2018; and G. A. Caceres et al. 2019) as well as manually vetting KOIs (J. J. Lissauer et al. 2024). In contrast, SysSim fits models to a selection of planets from the Kepler Data Release 25 (DR25) catalog. Specifically, SysSim focuses on KOIs designated candidates by the Kepler Robovetter that orbit FGK main-sequence stars, have an orbital period between 3 and 300 days, and have a planetary radius between $0.5R_\oplus$ and $10R_\oplus$ (M. Y. He et al. 2019).

The CDFs of the observed systems from L24 and the detected systems from SysSim are shown in Figure 4. When all 100 pregenerated catalogs are combined, the detected catalogs' distribution has a median gap complexity that is 0.1 larger than L24. Furthermore, 90% of the L24 systems have a gap complexity less than 0.53, while 90% of the detected SysSim systems have a gap complexity less than 0.71. The KS-test for these distributions results in $p = 0.00058$, showing a statistical difference. However, we show that the distribution of SysSim systems where only edge planets are undetected is closer to the actual observed distribution (KS-test $p$-value of 0.045). In contrast, SysSim adequately reproduces the mass partitioning and impact parameter dispersions of the observed L24 data set, as also shown in Figure 4. This result supports that SysSim underestimates the number of systems with highly uniform spacings, as previously discussed in M. Y. He et al. (2020), G. J. Gilbert & D. C. Fabrycky (2020), and E. V. Turtelboom et al. (2024). The SysSim maximum-

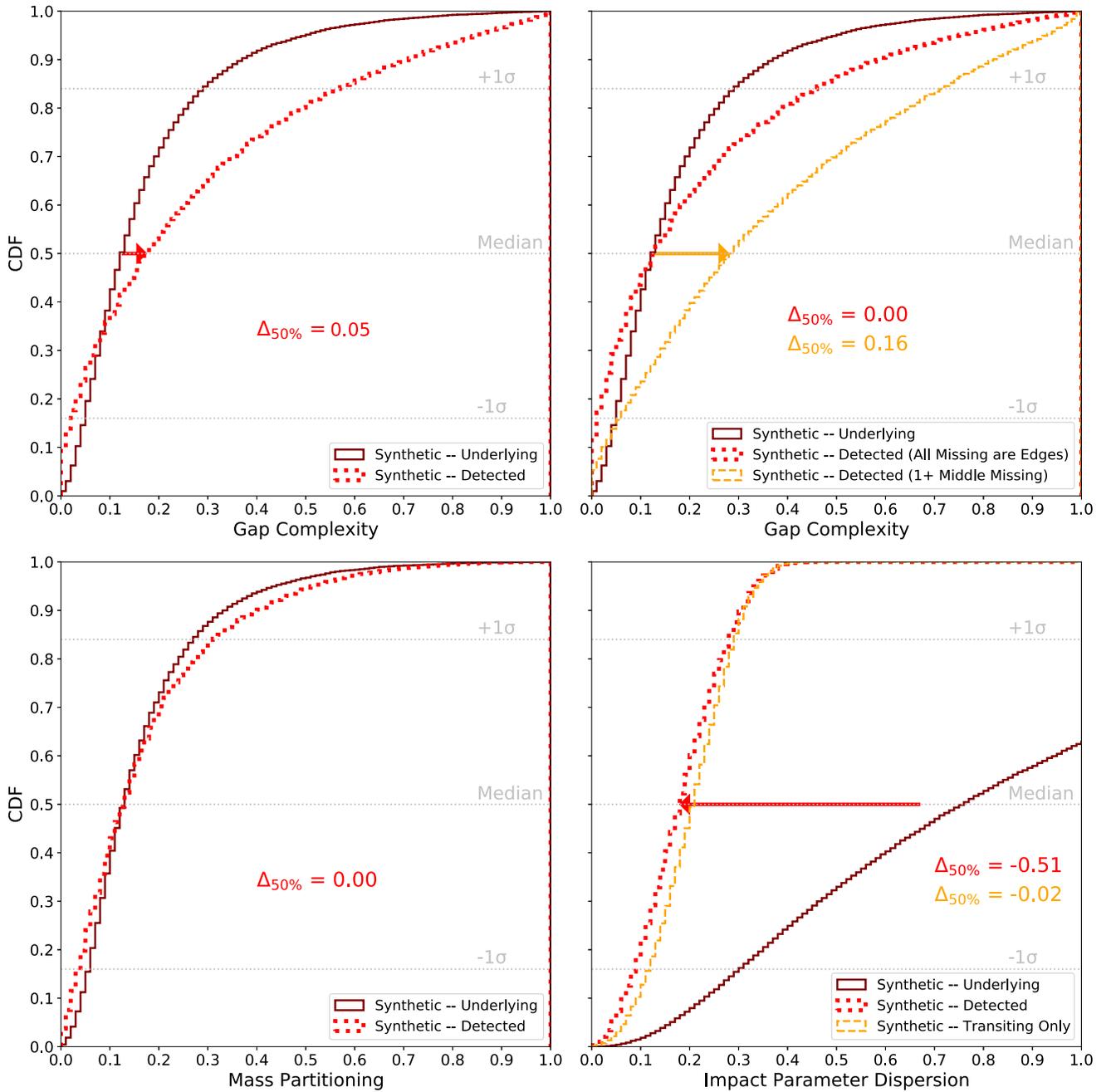

**Figure 3.** Cumulative distribution functions of gap complexity (top), mass partitioning (bottom left), and impact parameter dispersion (bottom right) for 100 combined `SysSim` synthetic planet catalogs. The solid lines (dark red) represent the underlying systems, while the dotted lines (red) represent the detected sample. In the top right panel, the detected sample is split between systems with all undetected planets being edge planets (red) and systems with at least one middle planet undetected (orange). In the bottom right panel, there is a third distribution (orange) that shows the impact parameter dispersion of only the transiting planets from the underlying catalog.

AMD model is not a perfect representation of the underlying distribution of planets, and thus, an underlying model that enforced more regular period spacing is needed to match the observed peas-in-a-pod pattern.

Furthermore, while the jackknife experiment on the L24 catalog and comparing the underlying and detected catalogs of `SysSim` both show similar results, caution must be used when extrapolating between the two scenarios. The knowledge gleaned from the `SysSim` experiment is helpful; however, the presence of potential missing planets in the L24 catalog cannot be extrapolated from the `SysSim` catalogs, as they are not a perfect reflection of the known Kepler planets. Thus, we must consider these as separate experiments.

### 2.4. The Role of Detection Bias, False Positives, and Imperfect Models

Could the apparent regular spacing of the peas-in-a-pod architecture be driven by Kepler's inherent detection biases, rather than astrophysical processes (e.g., W. Zhu 2020; also see L. Murchikova & S. Tremaine 2020)? If this were common, the discovery of additional (currently undetected) planets in multiplanet systems would increase their apparent uniformity in spacing, such that the median spacing diversity would decrease when removing planets. A potential scenario is the presence of an undetected smaller planet orbiting in between planets in a peas-in-a-pod system. If detection bias were the primary driver of the apparent uniform spacing, we would expect that systems would become more uniform when planets are missed. This would be reflected in our experiments if, when one or more planets are removed, the spacing and sizes would become more uniform. However, in both samples we tested (observed and synthetic), the opposite was true: removing one or more planets to simulate missing planets due to detection bias tended to increase the spacing diversity (reduce uniformity).

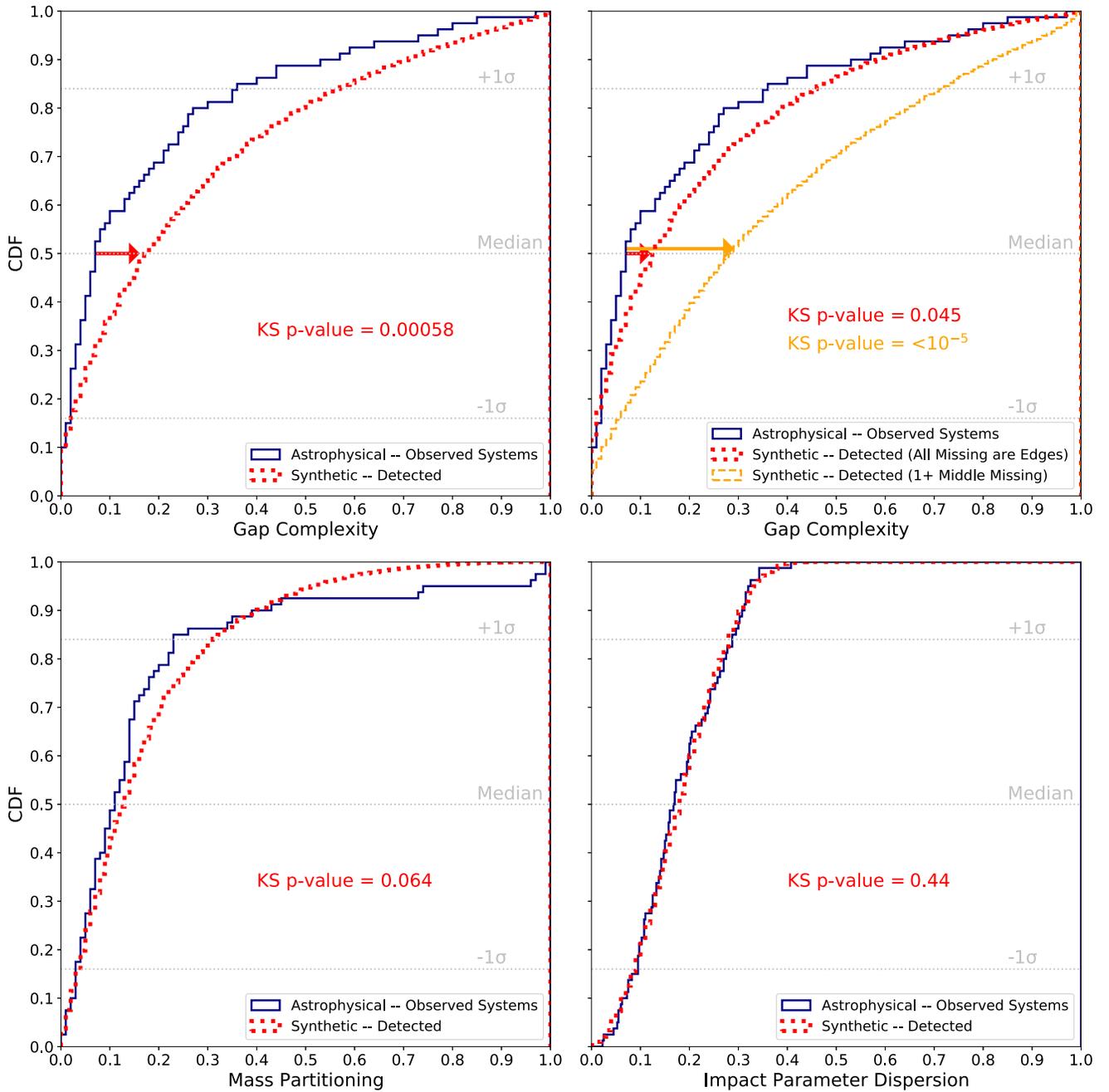

**Figure 4.** Comparing the cumulative distribution functions of gap complexity (top), mass partitioning (bottom left), and impact parameter dispersion (bottom right) for 80 4+ planet L24 systems (dark blue, solid) to 100 combined `SysSim` synthetic planet catalogs (red, dotted). In the top right panel, the detected sample is split between systems with all undetected planets being edge planets (red) and systems with at least one middle planet undetected (orange).

While there are indeed some systems for which removing one or more planets can decrease the gap complexity, this was not the norm, and so detection bias alone is insufficient to explain the regular spacing.

Could false positive identification of planets play a major role in the apparent commonality of peas-in-a-pod architectures? False positives are rare in multiplanet systems (J. J. Lissauer et al. [2012](), [2014]()). This is especially true in systems with a high multiplicity of transit-like signals. J. J. Lissauer et al. ([2014]()) provide a statistical framework to estimate the number of false positives in Kepler multiplanet systems. For their sample of 1054 planet candidates, they show that 99.8% should be real planets. This false alarm rate is so low that we expect our Astrophysical—Observed catalog to have <1 false positive. The synthetic systems do not contain false positives.

As G. E. P. Box (1976) famously said, "All models are wrong." Luckily, some models are useful, and each of the models we have considered in this Letter has a unique set of strengths and drawbacks. While the L24 sample[5] does not correct for detection biases, an advantage is that it directly uses the Kepler-observed high-multiplicity systems, and we are testing the effects of missing exactly one planet. The strength of the SysSim model is that we have a complete knowledge of the underlying architecture of every planetary system, but on the other hand, we know that the downselected Synthetic—Detected systems overpredict the number of systems with large gap complexity (see Section 2.3). Note that these models also lead to slightly different interpretations of the experiment: with L24, we tested the effect of missing one planet at a time, whereas with SysSim, we tested the effect of missing potentially multiple planets given Kepler's detection efficiency. These fairly different experiments lead to similar conclusions about how the metrics of system architectures can be biased.

### 3. Discussion and Summary

In this Letter, we conducted a suite of simple jackknife experiments in which we removed one or more planets at a time from proposed multiplanet systems. The goal of these experiments was to assess the extent to which "missing" (or failing to detect) a planet can bias system-wide architectural metrics. We have found:

1. In multitransiting systems, the failure to detect a planet (especially a middle planet) biases the gap complexity toward higher values.
2. The failure to detect a planet has essentially no effect on the measure of the mass partitioning.
3. The failure to detect a (transiting) planet has essentially no effect on the impact parameter dispersion.
4. Comparing the metrics of "detected" planets from the synthetic catalog to the observed planets, we found that the SysSim synthetic planets systematically:
   (a) Overestimate the gap complexity distribution (make systems where the spacing is too random).
   (b) Accurately predict the mass partitioning of planets.
   (c) Accurately predict the impact parameter distribution.
5. The apparent regular spacing of peas-in-a-pod is consistent with being astrophysical, rather than a result of detection biases.

These conclusions are as we expected for an underlying distribution of peas-in-a-pod: removing one planet out of a group of equally sized planets would not change the distribution of sizes of the group, while removing a planet from a group of equally spaced planets would lead to a noticeable gap in the architecture. If high gap complexity can point to the presence of undetected planets, three questions arise: "How many peas are we missing?," "How many more peas-in-a-pod systems will exist if we discover additional peas?," and "How do we detect these missing planets?"

We conclude that missing one or more planets tends to bias the gap complexity more than the other metrics. A synthetic population of planets that, after detection biases are applied, accurately reproduces the low gap complexity of the observed systems would be a welcome contribution.

### Acknowledgments

C.A.T., L.M.W., and M.Y.H. acknowledge support from the NASA Exoplanet Research Program (grant No. 80NSSC23K0269).

### ORCID iDs

C. Alexander Thomas  https://orcid.org/0009-0007-6386-151X
Lauren M. Weiss  https://orcid.org/0000-0002-3725-3058
Matthias Y. He  https://orcid.org/0000-0002-5223-7945

---

[5] For the purpose of this discussion, L24 is an empirical, nonparametric model.